\documentclass[prd,showpacs,letterpaper,nofootinbib,superscriptaddress,twocolumn]{revtex4}
\usepackage{dcolumn}
 \usepackage{graphicx}
 \usepackage{epstopdf}
 \usepackage{epsfig}
 \usepackage{amsmath,amssymb}
 \usepackage{natbib}

 \newcommand{\eV}{\mbox{ eV}}
 \newcommand{\kel}{\mbox{ K}}

  \newcommand{\iMpc}{\mbox{ Mpc$^{-1}$}}
    \newcommand{\MHz}{\mbox{ MHz}}

 \newcommand{\bq}{\begin{equation}}
 \newcommand{\eq}{\end{equation}}
 \newcommand{\bqa}{\begin{eqnarray}}
 \newcommand{\eqa}{\end{eqnarray}}
 \newcommand{\deriv}{{\rm d}}
  
  \newcommand{\la}{\lesssim}

 \newcommand{\lya}{Ly$\alpha$ }

 \newcommand{\apjl}{Astrophys. J. Lett.}
 \newcommand{\apjs}{Astrophys. J. Suppl. Ser.}
 \newcommand{\aap}{Astron. \& Astrophys.}
 \newcommand{\aj}{Astron. J.}
 \newcommand{\mnras}{Mon. Not. Roy. Astron. Soc.}
 \newcommand{\physrep}{Phys. Rep.}
 \newcommand{\araa}{Ann. Rev. Astron. \& Astrophys.}

\begin{document}
 
\title{Evolution of the 21 cm signal throughout cosmic history}

\author{Jonathan R.~Pritchard}\thanks{Hubble Fellow}
\email{jpritchard@cfa.harvard.edu}
\affiliation{Institute for Theory \& Computation, 
Harvard-Smithsonian Center for Astrophysics, 
60 Garden St., Cambridge, MA 02138}
\author{Abraham Loeb}
\affiliation{Institute for Theory \& Computation,
Harvard-Smithsonian Center for Astrophysics, 
60 Garden St., Cambridge, MA 02138}
 
 
 \begin{abstract}

The potential use of the redshifted 21 cm line from neutral hydrogen for
probing the epoch of reionization is motivating the construction of several
low-frequency interferometers. There is also much interest in the
possibility of constraining the initial conditions from inflation and the
nature of the dark matter and dark energy by probing the power-spectrum of
density perturbations in three dimensions and on smaller scales than probed
by the microwave background anisotropies.  Theoretical understanding of the
21 cm signal has been fragmented into different regimes of physical
interest.  In this paper, we make the first attempt to describe the full
redshift evolution of the 21 cm signal between $0<z<300$.

We include contributions to the 21 cm signal from fluctuations in the gas
density, temperature and neutral fraction, as well as the \lya flux, and
allow for a post-reionization signal from damped Ly$\alpha$ systems.  Our
comprehensive analysis provides a useful foundation for optimizing the
design of future arrays whose goal is to separate the particle physics from
the astrophysics, either by probing the peculiar velocity distortion of the
21 cm power spectrum, or by extending the 21 cm horizon to $z\gtrsim 25$
before the first galaxies had formed, or to $z\la 6$ when the residual
pockets of hydrogen trace large scale structure.

\end{abstract}
 
 
 
 \maketitle
 
 \section{Introduction} 
 \label{sec:intro}
 
Modern cosmology has advanced considerably in recent years due to precise
measurements of the cosmic microwave background (CMB) and large-scale
structure \citep{SDSS}.  These data sets probe the distribution of matter
at redshifts $z\sim 10^3$ and $z\la 0.3$, respectively, and only cover
$\sim 0.1\%$ of the entire comoving volume of the observable Universe
\citep{LW08}.  In addition, significant information about individual galaxies and quasars
out to $z=7$ has been obtained through the efforts of many telescopes at
multiple wavelengths \citep{Ellis}.  Nonetheless, our knowledge and
understanding of the evolution of most of the baryonic matter, which
resided in the intergalactic medium (IGM), is limited.  Constraints from
WMAP3 \citep{spergel2006} on the optical depth to the CMB
($\tau=0.093\pm0.029$) only weakly constrain the redshift of reionization.
The presence of a Gunn-Peterson trough \citep{gp1965} in quasar spectra at
$z>6.5$ provides a loose limit on the end of reionization \citep{Fan}.
Absorption by the \lya forest in quasar spectra constrain the IGM
temperature to be $T_K=2\times10^4\kel$ at $z\sim 4$ \citep{Hui}.  Beyond
these fragmentary data points lies a complicated story of ionization and
heating of the IGM going back to the formation of the first stars
\citep{Loeb,Loeb2}.

Low frequency observations of the redshifted 21 cm line of neutral hydrogen
present one of the most promising avenues for exploring the entirety of the
poorly constrained period between recombination and reionization
\citep{fob,BL}.  Additionally, mapping hydrogen throughout this period and
at lower redshifts raises the prospect of measuring primordial density
fluctuations over a much larger volume than that probed by the CMB or
galaxy redshift surveys \citep{loeb_zald2004,WL08}.  Several experiments
are currently being constructed (such as
MWA\footnote{http://www.haystack.mit.edu/ast/arrays/mwa/},
LOFAR\footnote{http://www.lofar.org/}, PAPER
\footnote{http://astro.berkeley.edu/$\sim$dbacker/EoR/},
21CMA\footnote{http://web.phys.cmu.edu/$\sim$past/}) and more ambitious
designs are being planned (SKA\footnote{http://www.skatelescope.org/})
to detect the theoretically-predicted 21 cm signal.

The volume to be mapped can be loosely divided into three parts: {\it
(i)} the {\it ``Dark Ages''} between the epoch of cosmic recombination and
the appearance of the first galaxies ($25\lesssim z\lesssim 10^3$); {\it
(ii)} the epoch of reionization (EoR) during which hydrogen was ionized
\footnote{However, exotic physics such as the decay of dark matter
\citep{furlanetto2006dm} or evaporation of primordial black holes might
alter this picture dramatically.}  by the UV radiation from galaxies
($6\lesssim z\lesssim 25$); and {\it (iii)} the post-reionization epoch
($z\lesssim 6$). The transition between the first two epochs, characterized
by the appearance of the first luminous sources, is a period of ``Cosmic
Twilight", during which the radiation fields of the sources begins to
affect the gas, heating and exciting the largely neutral inter-galactic
medium (IGM).  Eventually, the small ionized regions surrounding groups of
sources expand and percolate beginning the process of reionization.  After
reionization, the residual neutral hydrogen is found in dense clumps (such
as damped Ly$\alpha$ systems \citep{wolfe2005}), self-shielded from ionizing
radiation.

Precise measurements of the power-spectrum of 21 cm brightness fluctuations
could constrain the initial conditions from inflation \citep{loeb_zald2004}
(including deviations from scale-invariance or Gaussianity), as well as the
nature of the dark matter (including a measurement of the neutrino mass in
the range expected from atmospheric neutrino experiments \citep{LW08}) and
dark energy \citep{wyithe2007bao}.  The 21 cm fluctuations are expected to
simply trace the primordial power-spectrum of matter density perturbations
either before the first galaxies had formed ($z\gtrsim 25$)
\citep{loeb_zald2004,lewis2007} or after reionization ($z\lesssim 6$) --
when only dense pockets of self-shielded hydrogen survive
\citep{wyithe2007,Pen}.  During the epoch of reionization, the fluctuations
are mainly shaped by the topology of ionized regions
\citep{mcquinn2005,santos2007,Iliev}, and thus depend on astrophysical
details.  Nevertheless, even during this epoch, the line-of-sight
anisotropy of the 21 cm power spectrum due to peculiar velocities, can in
principle be used to separate the implications for fundamental physics from
the unknown details of the astrophysics \citep{bl2005sep,mcquinn2005}.

Recent work has explored the 21 cm signal that may arise during
reionization \citep{zfh2004freq,fzh2004} from variations in the \lya flux
\citep{bl2005detect,chuzhoy2006,pritchard2006}, X-ray heating
\citep{pritchard2007xray}, gas density \citep{loeb_zald2004,lewis2007} or
peculiar velocities \citep{bl2005sep,bharadwaj2004,mcquinn2005}.  In
addition, the post-reionization signal has been explored by
\citet{wyithe2007}.  In general, previous studies have focused on
individual regions of parameter space where only one of two of these
fluctuations needs to be considered and the rest are assumed to be
negligible.  Here we attempt for the first time to present a global picture
of the redshift evolution of the 21 cm signal, including all relevant
fluctuations.  The full history can be used to select particular redshift
(frequency) windows and observing strategies within which fundamental
physics would be optimally constrained.

In analyzing the full history of 21 cm fluctuations, we seek to address two
main issues.  First, we would like to identify the redshift at which the
presence of astrophysical sources begin to modify the pristine cosmological
signal.  This defines the minimum redshift that must be accessible for a
future observatory aiming to detect directly the pristine signal (in
analogy with CMB experiments).  Although the answer is model-dependent, we
will show that the rapidity with which structures form once non-linear
collapse gets underway means that the model dependence is weak.  The weak
dependence we find is generic to a Gaussian random field of initial density
perturbations, as expected from inflation, for which the fraction of mass
incorporated in collapsed objects grows exponentially (on the high-density
tail of the Gaussian) at early times.  Second, we wish to address the
possibility of using the angular separation proposed by \citet{bl2005sep}
to recover cosmological information from 21 cm observations after
astrophysical effects become large. We will show that there exists a window
in redshift where this technique might be used to extract the dark-matter
power spectrum despite the complication of other sources of fluctuations.
Finally, we would like to identify the primary astrophysical parameters
that control the 21 cm signal at subsequent epochs.

The paper is organized as follows.  In \S\ref{sec:theory} we discuss the
calculation of the 21 cm signal, attempting to bring together a variety of
different sources of fluctuations.  The results of these calculations are
illustrated in \S\ref{sec:results}.  In \S\ref{sec:observation}, we discuss
the experimental requirements for detecting the 21 cm signal and explore
the performance of three fiducial experiments.  Finally, we conclude in
\S\ref{sec:conclude} and discuss future prospects.

Throughout this paper, we assume a cosmology with $\Omega_m=0.26$,
$\Omega_\Lambda=0.74$, $\Omega_b=0.044$,
$H_0=100h\,\rm{km\,s^{-1}\,Mpc^{-1}}$ with $h=0.74$, $n_S=0.95$, and
$\sigma_8=0.8$, consistent with the latest measurements
\citep{spergel2006}.

 \section{Theoretical framework} 
 \label{sec:theory}
 
\subsection{21 cm brightness temperature}
We begin by briefly summarizing the physics of the 21 cm signal and refer
the interested reader to \citet*{fob} for further details.  The 21 cm line
of the hydrogen atom results from hyperfine splitting of the $1S$ ground
state due to the interaction of the magnetic moments of the proton and the
electron.  The HI spin temperature $T_S$ is defined through the ratio
between the number densities of hydrogen atoms in the $1S$ triplet and $1S$
singlet levels, $n_1/n_0=(g_1/g_0)\exp(-T_\star/T_S)$, where $(g_1/g_0)=3$
is the ratio of the spin degeneracy factors of the two levels, and
$T_\star\equiv hc/k\lambda_{21 \rm{cm}}=0.068\,\rm{K}$.  
The optical depth of this transition is small at all relevant redshifts,
yielding brightness temperature fluctuations
\begin{multline}\label{brightnessT}
T_b=27  x_{\rm{HI}}\left(1+\frac{4}{3}\delta_b\right)\left(\frac{\Omega_bh^2}{0.023}\right)\left(\frac{0.15}{\Omega_mh^2}\frac{1+z}{10}\right)^{1/2}\\ \times\left(\frac{T_S-T_\gamma}{T_S}\right)\,\rm{mK},
\end{multline}
Here $x_{\rm{HI}}$ is the neutral fraction of hydrogen, $\delta_b$ is the
fractional overdensity in baryons, and the factor of $4/3$ arises from
including the effect of peculiar velocities and averaging the signal over
angles.  The spin temperature is given by
\begin{equation}
T_S^{-1}=\frac{T_\gamma^{-1}+x_\alpha T_\alpha^{-1}+x_c T_K^{-1}}{1+x_\alpha+x_c},
\end{equation}
where $T_\alpha$ is the color temperature of the \lya radiation field at
the \lya frequency and is closely coupled to $T_K$ by recoil during
repeated scattering, and $x_c, x_\alpha$ are the coupling coefficients to
atomic collisions and Ly$\alpha$ photons, respectively.  The spin
temperature becomes strongly coupled to the gas temperature when
$x_{\rm{tot}}\equiv x_c+x_\alpha\gtrsim1$.

The collisional coupling coefficient is given by
\begin{equation}
x_c=\frac{4T_\star}{3A_{10}T_\gamma}\left[\kappa^{HH}_{1-0}(T_k)n_H +\kappa^{eH}_{1-0}(T_k)n_e\right],
\end{equation}
where $A_{10}=2.85\times10^{-15}\,\rm{s}^{-1}$ is the spontaneous emission
coefficient, $\kappa^{HH}_{1-0}$ is tabulated as a function of $T_k$
\citep{allison1969,zygelman2005} and $\kappa^{eH}_{1-0}$ is taken from
Ref. \citep{furlanettobros}. For a more detailed analysis of the collisional
coupling, see Ref. \citep{hirata2006col}.

The Wouthysen-Field effect \citep{wouth1952,field1958} coupling is given by
\begin{equation}\label{xalpha}
x_\alpha=\frac{16\pi^2T_\star e^2 f_\alpha}{27A_{10}T_\gamma m_e c}S_\alpha J_\alpha,
\end{equation}
where $f_\alpha=0.4162$ is the oscillator strength of the \lya
transition. $S_\alpha$ is a correction factor of order unity, which
describes the detailed structure of the photon distribution in the
neighborhood of the \lya resonance
\citep{chen2004,hirata2006lya,chuzhoy2006heat,furlanetto2006heat}.  We make
use of the approximation for $S_\alpha$ outlined in
Ref. \citep{furlanetto2006heat}.  For the models considered in this paper,
\lya coupling dominates over collisional coupling at redshifts
$z\lesssim25$, although we consider this more carefully later.

Fluctuations in the 21 cm signal may be expanded to linear order \citep{fob}
\begin{equation}\label{deltaTb}
\delta_{T_b}=\beta_b\delta_b+\beta_x\delta_x+\beta_\alpha\delta_\alpha+\beta_T\delta_T-\delta_{v},
\end{equation}
where each $\delta_i$ describes the fractional variation in the quantity $i$ and we include fluctuations in the baryon density (b), neutral fraction (x), \lya coupling coefficient ($\alpha$), gas temperature (T), and 
line-of-sight peculiar velocity gradient (v). The expansion coefficients are given by
\begin{eqnarray}
\beta_b&=&1+\frac{x_c}{x_{\rm{tot}}(1+x_{\rm{tot}})},\\
\beta_x&=&1+\frac{x_c^{HH}-x_c^{eH}}{x_{\rm{tot}}(1+x_{\rm{tot}})},\nonumber\\
\beta_\alpha&=&\frac{x_\alpha}{x_{\rm{tot}}(1+x_{\rm{tot}})},\nonumber\\
\beta_T&=&\frac{T_\gamma}{T_K-T\gamma}\nonumber\\
&+&\frac{1}{x_{\rm{tot}}(1+x_{\rm{tot}})}\left(x_c^{eH}\frac{\deriv \log\kappa_{10}^{eH}}{\deriv \log T_K}+x_c^{HH}\frac{\deriv \log\kappa_{10}^{HH}}{\deriv \log T_K}\right)\nonumber.
\end{eqnarray}
It is important to realise that since fluctuations in $x_H$ can be of order
unity, an expansion to second order becomes necessary near to the end of
reionization.  This leads to quartic terms contributing to the power
spectrum.

Noting that in Fourier space $\delta_{\partial v}=-\mu^2\delta$ \citep{bharadwaj2004}, where $\mu$ is the angle between the line of sight and the wavevector $\mathbf{k}$ of the Fourier mode, we may use equation \eqref{deltaTb} to form the power spectrum 
\begin{eqnarray}\label{Pexpansion}
P_{T_b}(k,\mu)&=&P_{bb}+P_{xx}+P_{\alpha\alpha}+P_{TT}+2P_{bx}\nonumber\\
&&+2P_{b\alpha}+2P_{bT}+2P_{x\alpha}+2P_{xT}+2P_{\alpha T}\nonumber\\
&&+P_{x\delta x\delta}+ {\rm other\,quartic\,terms}\nonumber\\
&+&2\mu^2\left(P_{b\delta}+P_{x\delta}+P_{\alpha\delta}+P_{T\delta}\right) \nonumber\\
&+&\mu^4P_{\delta\delta}\nonumber\\
&+&2P_{x\delta\delta_v x}+P_{x\delta_v\delta_v x}\nonumber\\ 
&&+{\rm other\,quartic\,terms\, with\, }\delta_v.
\end{eqnarray}
Here we note that all quartic terms must be quadratic in $x_H$ and separate them depending upon whether they contain powers of $\delta_v$ or not.  Those that contain powers of $\delta_v$ will not be isotropic and so contribute to the angular dependence of $P_{T_b}$ (see \citep{mcquinn2005} for further discussion).  

We may rewrite Eq. \eqref{Pexpansion} in more compact form
\begin{equation}
P_{T_b}(k,\mu)=P_{\mu^0}(k)+\mu^2P_{\mu^2}(k)+\mu^4 P_{\mu^4}(k)+P_{f(k,\mu)}(k,\mu),
\end{equation}
where we have grouped those quartic terms with anomalous $\mu$ dependence into the term $P_{f(k,\mu)}(k,\mu)$.  In principle, high precision measurements of the 3D power spectrum will allow the separation of $P_{T_b}(k,\mu)$ into these four terms by their angular dependence on powers of $\mu^2$ \citep{bl2005sep}.  The contribution of the $P_{f(k,\mu)}(k,\mu)$ term, with its more complicated angular dependence, threatens this decomposition \citep{mcquinn2005}.  Since this term is only important during the final stages of reionization, we will ignore it in this paper noting only that the angular decomposition by powers of $\mu^2$ may not be possible when ionization fluctuations are important.

It is unclear whether the first generation of 21 cm experiments will be able to achieve the high signal-to-noise required for this separation \citep{mcquinn2005}.  Instead, they might measure the angle averaged quantity
\begin{equation}
\bar{P}_{T_b}(k)=P_{\mu^0}(k)+P_{\mu^2}(k)/3+P_{\mu^4}(k)/5 ,
\end{equation}
(where we neglect the $P_{f(k,\mu)}(k,\mu)$ term).  In presenting our
results, we will concentrate on $\bar{P}_{T_b}(k)$, which is easiest to
observe and discuss the angular separation separately in
\S\ref{ssec:decomposition}.  We will typically plot the power per
logarithmic interval $\Delta=[k^3 P(k)/2\pi^2]^{1/2}$.

 \subsection{Astrophysical modeling}
Before calculating the evolution of the fluctuations, we must first provide an astrophysical model for the sources of radiation that modify the 21 cm signal.  We follow the model of \citet{furlanetto2006} with some minor modifications.
 
We must specify a luminosity and spectrum for sources of ionizing photons,
Lyman series photons, and X-rays.  The variation in spectra has relatively
little effect on the fluctuations (but see Refs. \citep{chuzhoy2006} and
\citep{pritchard2007xray} for further discussion on this point), so that
for our purposes we will fix the spectrum and consider models with
different luminosities.  We specify three key parameters for our
astrophysical sources: the number of ionizing photons produced per baryon
in stars that contribute to ionizing the IGM $N_{\rm ion,IGM}=f_{\rm esc}
N_{\rm ion}$, the number of \lya photons produced per baryon in stars
$N_\alpha$, and the energy in X-rays produced per baryon in stars
$\epsilon_X$.  With this parametrization, we are separating the physics of
the sources from the star formation history, as the quantity of physical
relevance will be the above numbers multiplied by the efficiency of star
formation $f_*$.  For example, the conventional definition of the ionizing
efficiency $\zeta=N_{\rm ion, IGM}f_*$, gives the important quantity when
tracking reionization.

For convenience, we further define $f_\alpha$ and $f_X$ via
$N_\alpha=f_\alpha N_{\rm \alpha,ref}$ and $\epsilon_X=f_X \epsilon_{X,
\rm{ref}}$ where we take the reference values appropriate for normal
(so-called, {\it Population II}) stars $N_{\rm \alpha,ref}=6590$ and for
starburst galaxies with a power-law spectrum of index $\alpha_S=1.5$ and
$\epsilon_{X, \rm{ref}}=560\eV$ \citep{glover2003}.  We use the spectra
appropriate for these sources throughout.  For comparison, in this
notation, the very massive ({\it Population III}) stars have
\citep{bromm2004}, $N_\alpha=3030$ ($f_\alpha=0.46$), when the
contribution from higher Lyman series photons is included.  We expect the
value of $f_\alpha$ to be close to unity.  In contrast, $f_X$ is relatively
unconstrained with values $f_X\lesssim10^3$ possible without violating the
WMAP CMB optical depth constraints.  Constraining this parameter will mark
a step forward in our understanding of the thermal history of the IGM and
the population of X-ray sources at high redshifts.

Once reionization is complete we assume that photo-ionization heats the gas to $3\times10^4 {\,\rm K}$.  This is needed only for determining the thermal cutoff scale for the baryonic fluctuations for the post-reionization 21 cm signal.

\subsection{Calculation of fluctuations}
The focus of this paper is to combine different sources of 21 cm
fluctuation into a complete picture of the evolution of the 21 cm signal
from the epoch when the cosmic gas thermally decoupled from the CMB ($z\sim
200$) to the present time.  In calculating the various sources of
fluctuations, we draw heavily upon the existing literature.  Below we
briefly discuss the relevant calculations and then describe how we piece
them together.

The 21 cm signal has been most extensively studied within the epoch of
reionization (EoR), where it is most accessible to observations.  During
this epoch large-scale (Mpc) ionized (HII) regions grow around clusters of
sources and delineate a density contrast of order unity in the hydrogen
abundance, far greater than the underlying fluctuations in the matter
density on the same scales \citep{bl2001}.  Analytic models of the HII
bubble size and resulting 21cm correlations were formulated by
\citet{fzh2004}, who considered bubbles to self-ionize once they contained
a sufficient mass fraction in galaxies (ionizing sources).  This basic
formalism has been applied to numerical simulations \citep{zahn2007} for
more detailed calculations of the ionization power spectrum.  Complementary
simulations by \citet{trac2007} have been compared directly to the
\citet{fzh2004} model showing reasonable agreement \citep{santos2007}.  For
computational ease, we choose to use the fully analytic calculation of
\citep{fzh2004} for the ionization fluctuations, including a correction for
an error in the bubble bias as noted by \citet{mcquinn2005ksz}.  This
prescription allows us to calculate $P_{xx}$ and $P_{x\delta}$.  The key
parameter in the \citet{fzh2004} model is the ionizing efficiency of
sources $\zeta$; we extract this redshift-dependent quantity from our
astrophysical model by relating the ionized fraction $x_i$ to the total
collapse fraction $f_{\rm coll}$ through $x_i=\zeta f_{\rm coll}$.
 
Fluctuations in the \lya coupling originate from spatial variation of the
\lya flux, which arise primarily due to the strong clustering of early
sources.  The calculation of these fluctuations has been discussed in
detail by \citet{bl2005detect} and \citet{pritchard2006}.  We consider
fluctuations from \lya photons arising from both stellar and X-ray sources
\citep{mmr1997,chen2006,chuzhoy2006}, which have different scale
dependence.  We follow the model of \citet{bl2005detect} in calculating
$\delta_\alpha$ by integrating the flux from all sources weighted
appropriately by distance and in addition account for higher Lyman series
photons \citep{pritchard2006,hirata2006lya}.  Recent investigations have
shown that including the scattering of \lya photons in the IGM near to
sources redistributes the \lya flux leading to greater power on large
scales \citep{semelin2007,chuzhoy2007,naoz2007}. We neglect this effect
estimating that this would change our results at the $10-20\%$ level when
\lya fluctuations dominate.
  
Temperature fluctuations arise initially through the competition between
adiabatic cooling and Compton scattering on the CMB (which drives the gas
towards being isothermal).  After thermal decoupling at $z\sim 200$, the
former process dominates and temperature fluctuations grow since the rate
of cooling is density dependent.  Once star-formation begins, X-ray heating
drives large temperature variation due to its highly inhomogeneous
nature.  We calculate the evolution of temperature fluctuations using the
formalism of \citet{pritchard2007xray} including all three
processes. Temperature fluctuations may also arise as a result of exotic
heating mechanisms such as the decay of dark matter
\citep{furlanetto2006dm}.  In this paper, we are interested in producing a
``standard" picture of the 21 cm signal and will leave exotic deviations to
future work.

Before star-formation, 21 cm fluctuations are determined by simple atomic
physics and so, potentially, provide a window into the physics of
inflation, via precise measurements of the spectrum of density fluctuations
\citep{loeb_zald2004}.  Detailed numerical work \citep{naoz2005,lewis2007}
has shown this period to be rich in physics as recombination, thermal
coupling, gas pressure and other physical processes interact to determine
the final 21 cm signal.  Since Compton drag on the CMB distinguishes
between baryons and dark matter, detailed calculation must track the
evolution of both components separately \citep{BLBAO}.  After thermal
decoupling, however, baryons fall into the potential wells created by the
dark matter and by $z\lesssim30$ it is a good approximation to take the
baryons as tracking the dark matter, although this ignores the presence of
acoustic oscillations in the baryon component and pressure smoothing on
small scales.  Baryon acoustic oscillations (BAO) represent a major target
for galaxy surveys and 21 cm observations since they help to constrain dark
energy via the angular-diameter-distance redshift relation
\citep{se2003probe}.  21 cm observations of baryon oscillations complement
galaxy surveys well \citep{BLBAO,wyithe2007bao,Pen}, since they cover a
higher redshift range and have a different set of biases than galaxy
surveys which may be affected by inhomogeneous galaxy formation
\citep{pritchard2007bub}.  

In our calculations, we assume that baryons trace the underlying density
field on large scales but are smoothed by the finite pressure of the gas on
small scales \citep{shapiro1994} so that $\delta_b=\delta (1+k^2
R_F^2)^{-1}$, where $R_F=1/k_F$ and $k_F$ is the filtering scale
\citep{gnedin2000}.  Thermal broadening introduces a second smoothing scale
$R_T$ \citep{bl2005detect}, which we incorporate through a Gaussian cutoff
factor, $\exp(-k^2 R_T^2)$.  Until X-ray heating becomes important
$R_F>R_T$, while afterwards $R_T>R_F$ as the filter mass takes a finite
period of time to respond to the greatly increased gas temperature.  These
smoothing scales are small (typically $R_T\sim R_F\sim{\rm kpc}$) until
photoionization heats the gas to $T_K\sim10^4\kel$ and smoothing becomes
important for wavenumbers $k\gtrsim10\iMpc$.  We calculate linear density
fluctuations using the fitting function of \citep{eisenstein1998}.
Non-linear density effects become important even at high redshift and we
incorporate these using the halo model \citep{cooray2002halo}.

After reionization is complete, a few percent of all neutral hydrogen (by
mass) remain in over dense clumps that are self-shielded
($\Omega_{HI}=10^{-3}$).  These clumps are observed as high-column density
absorbers in quasar spectra in the form of Lyman-limit or damped Ly$\alpha$
systems \citep{wolfe2005}.  Fluctuations in the 21 cm brightness from this
post-reionization epoch will trace the density fluctuations with a nearly
constant bias \citep{wyithe2007}.  Following the calculations of
\citet{wyithe2007}, we take the post-reionization signal to be
$P_{21}(k)=b_{21}^2P_{\delta\delta}(k)$ with $b_{21}=0.03\,{\rm K}$.
Although the fluctuations here are much weaker than before reionization,
the vastly diminished foreground at the appropriate higher frequencies
makes observations feasible.  Observation of fluctuations in this regime
may be used to detect baryon acoustic oscillations in the matter power
spectrum and constrain the equation of state of the dark energy at higher
redshifts than those accessible to supernova or galaxy surveys
\citep{wyithe2007bao}.

The existence of numerous cross-terms in Eq. \eqref{Pexpansion} connecting
different fluctuations requires some thought. Following the approach of
\citet{bl2005detect}, we relate $\delta_\alpha$ to the baryon density field
using a scale-dependent linear function
$\delta_\alpha=W_\alpha(k)\delta_b$.  Similarly we set
$\delta_T=g_T(k)\delta_b$ \citep{pritchard2007xray}.  In this way, we are
able to relate the cross-terms $P_{\alpha T}$, $P_{bT}$, $P_{b\alpha}$, and
equivalent $\delta$ terms to $P_{\delta\delta}$.  Since the \lya flux, gas
temperature, and baryon density do not interact directly with one another
these cross-terms should be a good approximation.

The treatment of the neutral fraction fluctuations is more complicated.
Once HII bubbles begin to form, we are faced with a two phase medium
composed of almost fully ionized HII regions and the largely neutral IGM
outside.  We must distinguish between $x_i$, the fraction of the cosmic gas
incorporated in fully ionized bubbles, and $x_e$ the ionized fraction in
the largely neutral IGM (which could be increased beyond its primordial
value due to diffuse X-rays).  Neutral fraction fluctuations arise from
both, although in most models that we consider $x_e$ remains small at all
times and so fluctuations in $x_e$ are also small.  The calculations of
$\delta_\alpha$ and $\delta_T$ by \citet{bl2005detect} and
\citet{pritchard2007xray} assume that the filling fraction of HII regions
is negligible. However, here we must relax that assumption in order to
calculate terms like $P_{x\alpha}$.  Since there is essentially no 21 cm
signal from the ionized HII regions, the HII regions act to mask the
fluctuations in $\alpha$ and T.  Overdense regions will host sources of
\lya and X-ray photons, as well as ionizing photons, suggesting that
cross-terms of the form $P_{x\alpha}$ and $P_{x T}$ may be complicated.  We
show later that these cross-terms are subdominant, so that, for a first
approximation, we may neglect these complications and model the cross-terms
as $P_{x\alpha}=W_\alpha(k)P_{x b}$, $P_{x T}=g_T(k) P_{x b}$, and so on.
Detailed calculation will require numerical simulation of the ionization
and radiation fields to determine the cross-terms.

We apply the model of \citet{fzh2004} to calculate $P_{xx}$, $P_{x\delta}$, and $P_{x\delta x\delta}$.  Note that this model first calculates the correlation functions and Fourier transforms.  For the quartic term, we ignore the possible connected part so that $P_{x\delta x\delta}=P_{xx}P_{\delta\delta}+2P_{x\delta}^2$.  Quartic terms involving $\delta_v$ contain angular dependence that threatens the possibility of the $\mu^2$ decomposition \citep{mcquinn2005}. The other quartic terms we handle in the same way as terms like $P_{x\alpha}$, first reducing them to a quartic of $\delta$ and X multiplied by factors of $W_\alpha$ and $g_T$, then calculating them in the above way.  These are generally small corrections, relevant only at the end of reionization.


\section{Results}
\label{sec:results}

\subsection{Mean History}
\begin{figure}[htbp]
\begin{center}
\includegraphics[scale=0.4]{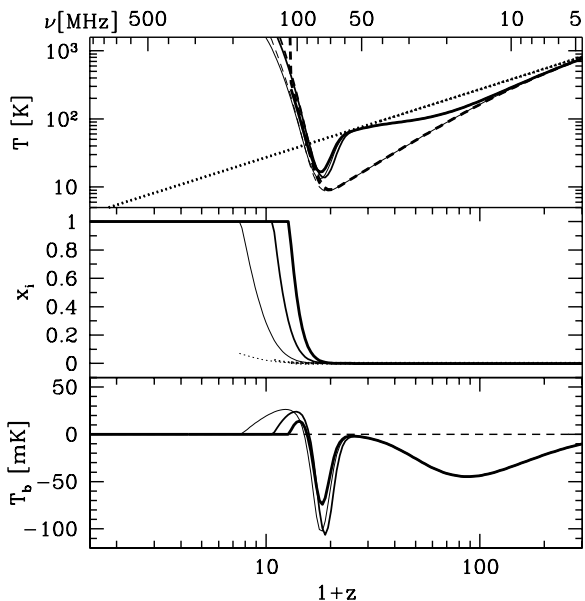}
\caption{{\em Top panel: }Evolution of the CMB temperature $T_{\rm CMB}$
(dotted curve),the gas kinetic temperature $T_K$ (dashed curve), and the
spin temperature $T_S$ (solid curve).  {\em Middle panel: }Evolution of the
gas fraction in ionized regions $x_i$ (solid curve) and the ionized
fraction outside these regions (due to diffuse X-rays) $x_e$ (dotted
curve). {\em Bottom panel: } Evolution of mean 21 cm brightness temperature
$T_b$.  In each panel we plot curves for model A (thin curves), model B
(medium curves), and model C (thick curves).}
\label{fig:tempZlong}
\end{center}
\end{figure}

We choose three astrophysical models chosen to reproduce the WMAP3 central
and $\pm1-\sigma$ for the cumulative electron-scattering optical depth
$\tau$.  Model A uses ($N_{\rm ion,IGM}$, $f_\alpha$, $f_X$, $f_*$) =
(200,1,1,0.1) giving $z_{\rm reion}=6.47$ and $\tau=0.063$. Model B uses
($N_{\rm ion,IGM}$, $f_\alpha$, $f_X$, $f_*$) = (600,1,1,0.2) giving
$z_{\rm reion}=9.76$ and $\tau=0.094$.  Model C uses ($N_{\rm ion,IGM}$,
$f_\alpha$, $f_X$, $f_*$) = (3000,0.46,1,0.15) giving $z_{\rm reion}=11.76$
and $\tau=0.115$.  The evolution of the mean temperatures with redshift for
these models is shown in Figure \ref{fig:tempZlong}.  The top axis shows
the relevant observing frequency range given by $\nu_{\rm obs}=\nu_{\rm
21cm}/(1+z)$.  Since we are leaving the cosmology unchanged between models,
the mean history only varies once star-formation becomes important around
$z\approx25$.  The most significant difference between these models is the
redshift of reionization that cuts off the 21 cm signal at low redshift and
is most responsible for the different $\tau$ values.  Model C has a
significantly decreased \lya flux, which shows itself through a decrease in
the strength of the absorption signal.  

The global evolution of the 21 cm signal is itself the target of several experiments, e.g. the ``Cosmological Reionization Experiment" (CoRE) \citep{chippendale2005} and the ``Experiment to Detect the Global EOR Signature" (EDGES) \citep{bowman2007edges}.  Although such an observation is conceptually simple, it is experimentally challenging.  Distinguishing between a global signal and other sources of all sky emission, including Galactic synchrotron, free-free radiation, and the CMB is difficult.  Experimental detection relies upon reionization proceeding rapidly leading to a distinctive step-like feature in the frequency direction, which would not be expected to be produced by the spectrally-smooth foregrounds.  With the assumption of sharp reionization, EDGES \citep{bowman2007edges} places an initial constraint that $\bar{T}_b<450 \,{\rm mK}$ at $z=8$.  While this is far from the expected signal amplitude, such constraints will improve with time.  Efforts are also underway to extend the frequency coverage to $\nu\approx50\, {\rm MHz}$ to access the transition from an absorption to emission signal.  As Figure \ref{fig:tempZlong} indicates, this transition is likely to be significantly larger in amplitude ($\sim100\,{\rm mK}$) than that at the end of reionization ($\sim20\,{\rm mK}$).

\subsection{Fluctuation History}
\label{ssec:f_history}

\begin{figure}[htbp]
\begin{center}
\includegraphics[scale=0.4]{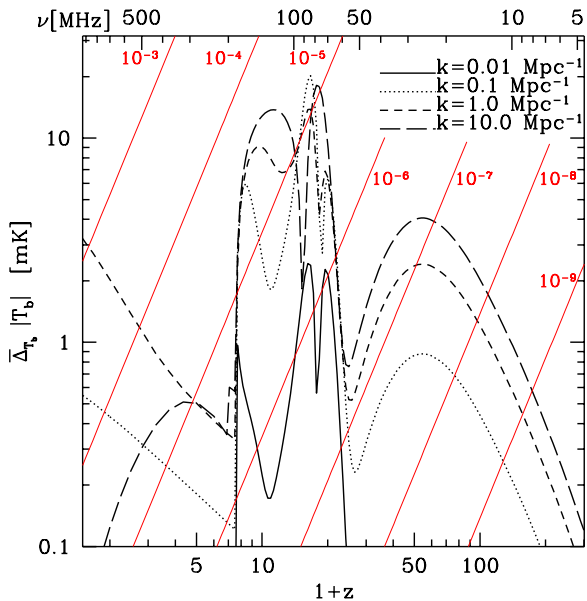}
\caption{Redshift evolution of the angle-averaged 21 cm power spectrum $\bar{\Delta}_{T_b}$ for Model A at $k=0.01$ (solid curve), 0.1 (dotted curve), 1.0 (short dashed curve), and 10.0 (long dashed curve)$\iMpc$. Reionization at $z=6.5$.}
\label{fig:powerZlong_low}
\end{center}
\end{figure}

\begin{figure}[htbp]
\begin{center}
\includegraphics[scale=0.4]{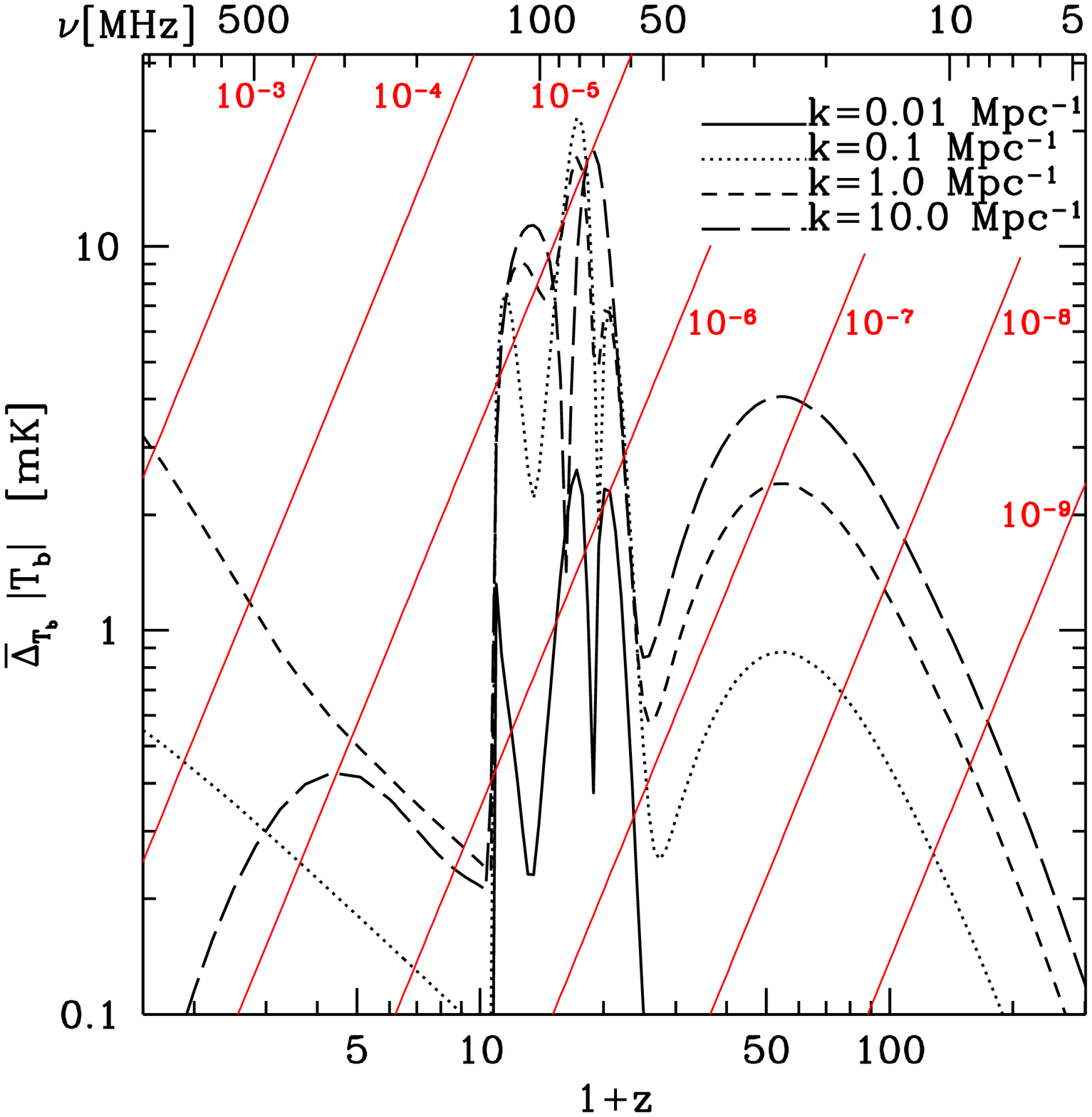}
\caption{Redshift evolution of the angle-averaged 21 cm power spectrum $\bar{\Delta}_{T_b}$ for Model B. Reionization at $z=9.8$.  Same line conventions as Figure \ref{fig:powerZlong_low}.}
\label{fig:powerZlong_mid}
\end{center}
\end{figure}

\begin{figure}[htbp]
\begin{center}
\includegraphics[scale=0.4]{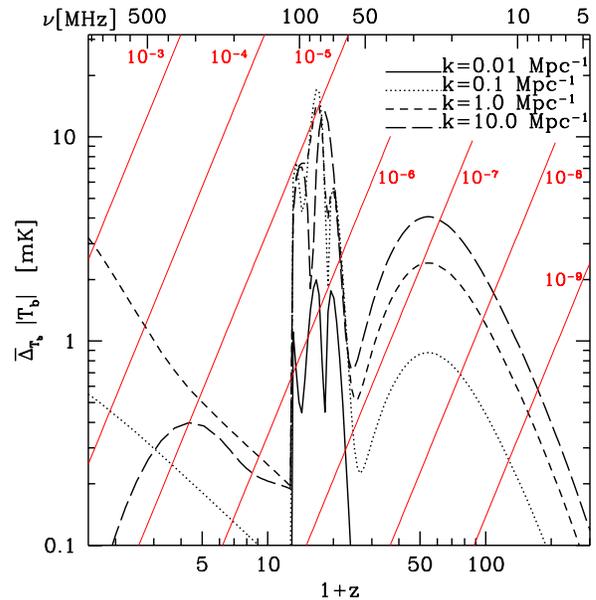}
\caption{Redshift evolution of the angle-averaged 21 cm power spectrum $\bar{\Delta}_{T_b}$ for Model C. Reionization at $z=11.8$.  Same line conventions as Figure \ref{fig:powerZlong_low}.}
\label{fig:powerZlong_high}
\end{center}
\end{figure}

The three dimensional nature of the 21 cm signal makes it difficult to
convey the evolution of the fluctuations with a single 2-dimensional plot.
We therefore plot the evolution of four individual comoving wavenumbers
$k=0.01$, 0.1, 1, and 10 $\iMpc$, spanning the range that might be
observed.  On large scales we expect contamination from foregrounds to
limit the detection of the power spectrum.  On small scales thermal
broadening of the 21 cm line will smooth the signal.  It is also to be
expected that many of our approximations will break down as small scale
information about the sources becomes important (see for example
\citep{lidz2007small} for the importance of higher order correlations on
small scales during reionization).  For the mean histories shown in Figure
\ref{fig:tempZlong}, we calculate the evolution of the 21 cm angle-averaged
power spectrum, which is plotted in Figures \ref{fig:powerZlong_low}, 
\ref{fig:powerZlong_mid}, and \ref{fig:powerZlong_high}, for models
A, B, and C, respectively.

The evolution of $\bar{\Delta}_{T_b}$ clearly shows three regimes: the
post-reionization regime at low redshifts ($z<z_{\rm reion}$) where the
21cm fluctuations from residual hydrogen follow the matter power spectrum,
an intermediate redshift regime ($x_{\rm reion}<z<z_{\rm trans}$) where
\lya coupling produces a large signal and complicated astrophysics leads to
significant scale dependence, and a high redshift collisionally-coupled
regime where 21 cm fluctuations track the density field ($z>z_{\rm
trans}\approx23$).  For pedagogical purposes, let us describe the evolution
on a single comoving scale (say, $k=0.1\iMpc$) and draw attention to the
main features.  Thermal decoupling at $z\sim200$ is a gradual process and,
initially, $\bar{\Delta}_{T_b}$ grows due to a combination of the growth of
density fluctuations and the steady gas cooling below $T_{\rm CMB}$.  As
the gas rarifies and cools, collisional coupling becomes less effective
and, at $z\sim60$, $\bar{\Delta}_{T_b}$ begins to decrease in amplitude.
Note that the continuing growth of structure offsets the turnover on
$\bar{\Delta}_{T_b}$ from the minimum of $T_b$, seen in Figure
\ref{fig:tempZlong} to occur at $z\approx90$.  As collisional coupling
diminishes, the signal drops towards zero.  This occurs while $T_K<30$, a
regime where $\kappa_{1-0}(T_K)$ drops exponentially with $T_K$
\citep{zygelman2005} and results in a rapid drop of the signal at
$z\lesssim 40$.  Before the signal drops all the way to zero, significant
star-formation occurs and the resultant \lya production leads to the
beginning of \lya coupling by $z\approx25$.  The exponential increase in
the global star formation rate at these redshifts is responsible for the
rapid increase in $T_b$ and $\bar{\Delta}_{T_b}$.  With this rise
in signal, we enter into a regime dominated by astrophysics and begin to
see complicated scale dependence.

Initially, \lya fluctuations boost the signal somewhat above the level of
density fluctuations alone.  However, X-ray heating follows not far behind
and contributes to $\bar{\Delta}_{T_b}$ with the opposite sign (hotter
regions produce a weaker absorption signal, see \citet{pritchard2007xray}).
In this competition, X-ray driven temperature fluctuations dominate causing
$\bar{\Delta}_{T_b}$ to pass through a zero point (seen as a sharp dip at
$z\sim18$ in all three plots).  Temperature fluctuations dominate as
$\bar{T}_K$ approaches $T_{\rm CMB}$ and $\bar{T}_b$ vanishes.  In
proceeding to the emission regime, we note based on Figure
\ref{fig:tempZlong} that the brightness fluctuations emitted $T_b$ are
generically smaller than they were during the absorption regime, leading to
a decreasing trend in $\bar{\Delta}_{T_b}$.  As reionization gets underway,
ionization initially causes $\bar{\Delta}_{T_b}$ to drop leading to a
pronounced dip in its evolution.  This occurs as a result of the clustering
of ionizing sources in over dense regions causing the ionized HII regions
to ``mask out" those dense regions that have been producing the strongest
21 cm signal.  As reionization proceeds, the contrast between ionized and
neutral regions comes to dominate and $\bar{\Delta}_{T_b}$, rises until
$x_H\sim0.5$ after which the contrast begins to drop.

Towards the end of reionization the signal drops sharply as very little gas
is left neutral.  The post-reionization signal grows slowly as the density
field grows.  Since by this time the gas is photo-heated to
$T_K\approx 30,000 \kel$ the thermal width of the 21 cm line is sufficient
to smooth out the signal on wavenumbers $k\gtrsim10\iMpc$.  This cutoff
potentially acts as a thermometer of the gas after reionization giving
information about the temperature of gas contained in dense clumps.

As a result of the interplay between the radiation fields, as
$\bar{\Delta}_{T_b}$ evolves it shows three peaks within the
astrophysics-dominated regime.  An important feature of this complicated
evolution is that the maximum amplitude of $\bar{\Delta}_{T_b}$ occurs at
different $k$ values for different redshifts.  Accurate observation and
modeling of this complicated evolution may provide important information
about the early radiation fields.
 

It is helpful to get a sense of how the amplitude of the signal compares
with galactic foregrounds.  We take the sky noise to be $T_{\rm sky}\approx
180\kel (\nu/180{\,\rm MHz})^{-2.6}$ (appropriate for galactic synchrotron
emission \citep{fob}), noting that the normalization depends upon the
region of sky being surveyed.  In Figure \ref{fig:powerZlong_low}, Figure
\ref{fig:powerZlong_mid}, and Figure \ref{fig:powerZlong_high} we plot $r
T_{\rm sky}(\nu)$ where $r$ ranges from $10^{-4}-10^{-9}$.  We see that
reducing foregrounds by a factor of $\sim10^{-5}$ is required to observe
fluctuations during reionization and cosmic twilight.  The difficulty
increases if reionization occurs early, which has the effect of compressing
the signal at high redshifts (model C).  The signal from astrophysics in
these three models begins at $\nu\approx60\MHz$ and continues to
$\nu\approx150\MHz$ although this upper limit is very sensitive 
to the details of reionization.

Removing foregrounds to the rather low level of $\sim10^{-7}$ is required
to have a hope of observing the {\it Dark Ages} in the range $z=30$--$50$
($25\MHz\lesssim\nu\lesssim60\MHz$).  Some hope is restored by noticing
that the rapid evolution of $\bar{\Delta}_{T_b}$ between $z=30-50$ mimics
the strong frequency dependence of the foregrounds.  If one can remove
foregrounds to the level needed to observe $z=30$, then the same level of
foreground subtraction would allow $z=50$ to be observed.  Beyond $z=50$,
the amplitude of $\bar{\Delta}_{T_b}$ decreases rapidly and becomes
progressively more difficult to observe.  Modifying the thermal history
between $z=30$ and recombination by introducing exotic physics can greatly
change this high redshift behavior.  Energy injection by decaying dark
matter \citep{furlanetto2006dm} can increase the residual electron fraction
in the IGM while maintaining collisional coupling to lower redshifts
boosting the high redshift 21 cm signal and improving the chance of
observing the {\it ``Dark Ages"}.

\subsection{Decomposition of fluctuations}
\label{ssec:decomposition}

The total 21 cm signal is composed of fluctuations in five quantities.
Thus, when we form the power spectrum from Equation \eqref{Pexpansion} we
must calculate 15 distinct correlations and cross-correlations.  In
addition, as fluctuations in the neutral fraction can be of order unity,
there are a number of fourth order correlations which must be calculated.
Separating out these different components will be difficult.  A first step
will be to exploit the angular dependence of the peculiar velocity term to
separate out the contributions from $P_{\mu^0}$, $P_{\mu^2}$, and
$P_{\mu^4}$.  Since each of these three terms depends on different
combinations of terms, measuring the three separately gives extra
discriminating power \citep{bl2005sep}.  This decomposition is complicated
by the presence of the $P_{xx\delta_b\delta_v}$ and
$P_{xx\delta_v\delta_v}$ terms, which show very different $\mu$ dependence
\citep{mcquinn2005}.  Since these terms will be large only once
reionization is well underway, we shall ignore the angular dependence of
these terms.

\begin{figure}[htbp]
\begin{center}
\includegraphics[scale=0.4]{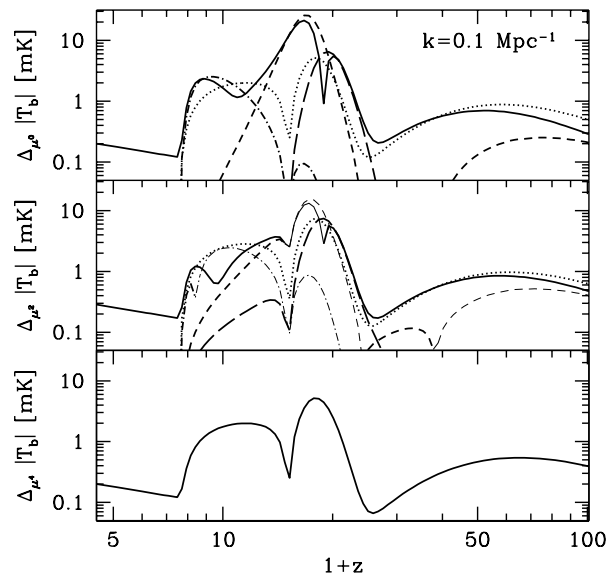}
\caption{Angular decomposition of 21 cm power spectrum for model A at
wavenumber $k=0.1\iMpc$.  {\em Top panel: }$\Delta_{\mu^0}(k)$ (solid
curve).  We also show the separate contribution from the {\it bb} (dotted
curve), $\alpha\alpha$ (long dashed curve), TT (short dashed curve), and XX
(dot-dashed curve) terms. Note that there are additional cross-terms, many
of which contribute to the power spectrum with negative sign, not shown
here.  {\em Middle panel: }$\Delta_{\mu^2}(k)$ (solid curve).  We also plot
the contribution from the bV (dotted curve), $\alpha$V (long dashed curve),
TV (short dashed curve), and XV (dot-dashed curve) terms, indicating the
sign of the contribution to the power spectrum as positive (thick curves)
or negative (thin curves).  {\em Bottom panel: }$\Delta_{\mu^4}(k)$ (solid
curve). }
\label{fig:muplot}
\end{center}
\end{figure}

We plot the different angular power spectra for model A in Figure \ref{fig:muplot}.  The $\mu^4$ term is the simplest, arising solely from the VV term, so that it tracks $P_{\delta\delta}$, making it ideal for trying to measure cosmology.  Its evolution is modulated by the evolution of the mean $T_b$, being divided into two absorption bumps and an emission bump.

The $\mu^0$ contribution contains contributions from 10 different terms,
and so we have not attempted to show its full decomposition. Instead, we
plot the contribution only from the auto-correlations, which gives some
idea of where different contributions are important.  Note that the
contribution of cross-terms is significant and that many cross-terms have
negative signs over some range of redshifts.  Hence, $\Delta_{\mu^0}$ can
be smaller than the sum of the auto-correlations would indicate.  Since the
$\mu^0$ term dominates $\bar{\Delta}_{T_b}$, its behavior is very similar
in form to that plotted in Figure \ref{fig:powerZlong_low}.  Temperature,
\lya, and ionization show clear overlapping regions of contribution.

Finally, the $\mu^2$ contains contributions from the bV, XV, $\alpha$V, and
TV cross-terms.  We have plotted these in the middle panel of Figure
\ref{fig:muplot} to illustrate where different effects become important.
Since these terms differ only in one element of the cross-correlation,
$\Delta_{\mu^2}$ is more sensitive to the form of the different
fluctuations than $\Delta_{\mu^0}$.  At $z\gtrsim22$, the dominant
contribution to $\Delta_{\mu^2}$ comes from the bV term although the TV
term is also significant.  The TV term changes sign at $z\approx38$ owing
to the behavior of $\beta_T$.  Below $z\approx22$ astrophysics becomes
important and the $\alpha$V term dominates for a short while before the TV
term overtakes it, causing $\Delta_{\mu^2}$ to change sign.  Around
$z\approx15$ all components drop towards zero briefly as $\bar{T}_b=0$.
This imprints an extra dip in the $\Delta_{\mu^2}$ power spectrum that is
not seen in $\Delta_{\mu^0}$, where the TT term dominates at this point.
Utilizing the different redshift evolution of these power spectra is likely
to greatly increase the chance of separating different astrophysical
sources of fluctuations.

\subsection{Separation of fundamental physics from astrophysics at high redshifts}
\label{ssec:transitionz}
We next turn our attention to characterizing the redshift at which the 21
cm signal becomes affected by the astrophysics of star formation.
Although, in principle, this could occur when any of $\beta_T\delta_T$,
$\beta_x\delta_x$, or $\beta_\alpha\delta_\alpha$ become large in
comparison with $\beta_b\delta_b$, as we have seen in the preceding
section, the onset of \lya coupling is likely to provide the first
astrophysical modification of the signal.  In order to quantify this
transition, we will consider the redshift $z_{\rm trans,\alpha}$ at which
$\beta_\alpha\delta_\alpha=\beta_b\delta_b$ for the first time.  Although
cosmological parameters may be extracted from components other than the
{\it bb}-term, this is the dominant component at high redshift.

 \begin{figure}[htbp]
\begin{center}
\includegraphics[scale=0.4]{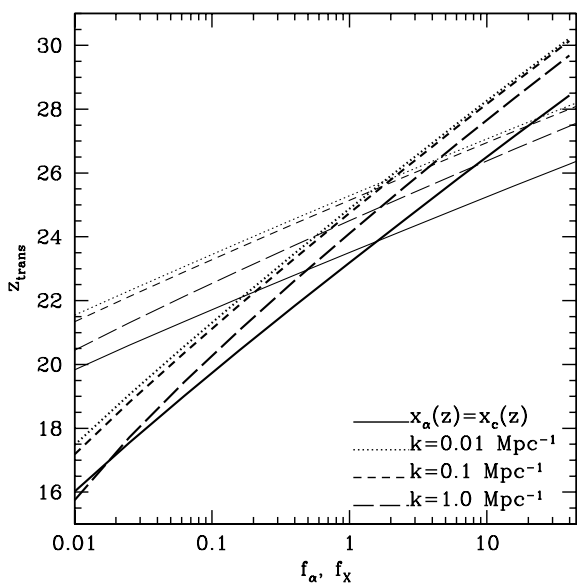}
\caption{Evolution of $z_{\rm trans,\alpha}$ with $f_\alpha$ (thin curves) and $f_X$ (thick curves) for $k=0.01$ (dotted curves), 0.1 (short dashed curve), and 1 $\iMpc$ (long dashed curve).  Also plotted is the redshift at which $x_\alpha=x_c$ (solid curves).}
\label{fig:transitionZ_nlya}
\end{center}
\end{figure}
Figure \ref{fig:transitionZ_nlya} shows the dependence of $z_{\rm
trans,\alpha}$ on the parameters $f_X$ and $f_\alpha$.  In order to
separate out the different effects, we consider two artificial situations.
First, we vary $f_\alpha$ using the fiducial value $f_X=1$, but neglecting
\lya photons produced by X-rays.  Next, we vary $f_X$ and neglect the
stellar \lya component, i.e. $f_\alpha=0$.  Thus, the plotted values of
$z_{\rm trans,\alpha}$ should be regarded as lower limits, since including
the \lya flux of the missing component will cause the transition to occur
slightly earlier.  All other parameters are set to those of model A.  We
plot $z_{\rm trans,\alpha}$ for three different wavenumbers, $k=0.01$, 0.1,
and 1$\iMpc$, noting that there is little difference between the three
curves.  For comparison, we also plot the redshift at which \lya and
collisional coupling are comparable, defined by $x_\alpha=x_c$.

The basic conclusion is that for a Gaussian field of initial density
perturbations, $z_{\rm trans,\alpha}$ depends only logarithmically on $f_x$
and $f_\alpha$.  This dependence arises because the fraction of mass in
galaxies grows exponentially with cosmic time on the high-density tail of
the Gaussian.  Exploring four orders of magnitude in $f_\alpha$ causes
$z_{\rm trans,\alpha}$ to vary by a redshift interval of $\Delta z=8$
($\Delta\nu\approx20\MHz$).  The same variation in $f_X$ causes $z_{\rm
trans,\alpha}$ to vary by $\Delta z=14$ ($\Delta\nu\approx40\MHz$).  This
extra dependence arises because modifying $f_X$ also changes $T_K$.  Note
that we can relate the uncertainty in the star formation efficiency $f_*$
to these results, provided we remember that changing $f_*$ will modify
the production of both UV and X-ray photons.

We can understand this parameter dependence, by considering the redshift at
which \lya coupling first becomes important.  Obtaining $x_\alpha=1$
requires a flux $J_\alpha=0.0767\left({1+z}/{20}\right)^2$ \lya photons
per baryon.  We can convert this into the fraction of baryons in stars
needed for \lya coupling
\begin{equation}\label{etaA}
\eta_\alpha\equiv \frac{J_\alpha}{N_\alpha}=8.2\times10^{-6}\left(\frac{1+z}{20}\right)^{-2}\left(\frac{6950}{N_\alpha}\right).
\end{equation}
To a good approximation, we can find the redshift of transition by setting
$\eta_\alpha=f_{\rm coll}f_*$ to find the point when enough baryons are in
stars for the transition to occur.  This will slightly overestimate the
redshift of transition, as we are not accounting for photons that redshift
out of the \lya resonance.  Since $f_{\rm coll}$ initially grows
exponentially with redshift we obtain a logarithmic dependence
on $f_*N_\alpha$.  A similar result will apply for \lya photons from X-ray
sources.

\begin{figure}[htbp]
\begin{center}
\includegraphics[scale=0.4]{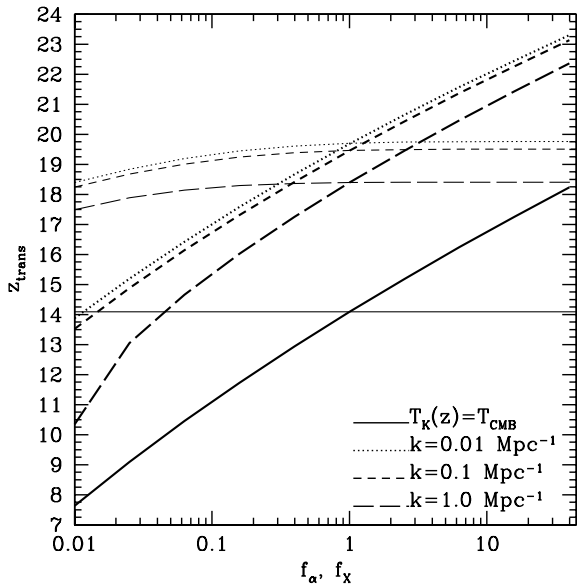}
\caption{Evolution of $z_{\rm trans,T}$ with the same line conventions as Figure \ref{fig:transitionZ_nlya}.  Also plotted is the redshift at which $T_K=T_{\rm CMB}$ (solid curves).}
\label{fig:transitionZ_temp}
\end{center}
\end{figure}

In more extreme astrophysical models, it is conceivable that intense X-ray
heating might affect the 21 cm signal before \lya coupling became
important.  We therefore also consider the redshift $z_{\rm trans,T}$ at
which $\beta_T\delta_T=\beta_b\delta_b$ for the first time.  This is
plotted in Figure \ref{fig:transitionZ_temp} for the same models used
above.  For comparison, we also plot the redshift at which the gas is
heated to $T_K=T_{\rm CMB}$, which occurs significantly later and marks the
transition from absorption to emission.  Unsurprisingly, $z_{\rm trans,T}$
shows very little dependence on $f_\alpha$.  The dependence on $f_X$ is
more pronounced.  In no part of the parameter space explored does $z_{\rm
trans,T}$ exceed $z_{\rm trans,\alpha}$, suggesting that \lya coupling will
always be the first source of astrophysical fluctuations. This need not be
the case if gas heating does not also produce \lya photons, for example if
shock heating dominates in the early Universe \citep{furlanetto2004}.  We
also note that values of $f_*f_X\gtrsim0.01$ are required to ensure that
$T_K>T_{\rm CMB}$ in the redshift range $z<12$ that will be probed by the
first low-frequency observatories.

In the same fashion as the onset of \lya coupling, we can seek to
understand the requirement for heating the gas above the CMB.  Assuming
that $T_K\ll T_{\rm CMB}$ at the onset of heating, an energy of $\Delta
E\approx 3k_B T_{\rm CMB}/2$ per baryon is required.  Using parameters
appropriate for X-ray emission for starburst galaxies and assuming that
$x_e$ is approximately the primordial value, yields the fraction of baryons
in stars required to heat the gas
\begin{equation}\label{etaT}
\eta_T\equiv \frac{\Delta E}{\epsilon_X}=1.15\times 10^{-4}\left(\frac{1+z}{20}\right)\left(\frac{560 \eV}{\epsilon_X}\right)\left(\frac{0.073}{f_{\rm heat}}\right).
\end{equation}
Comparison with Eq. \eqref{etaA} shows that heating the IGM to $T_{\rm
CMB}$ with X-rays requires considerably more star formation than producing
\lya coupling.  While generalizing this argument rigorously from the mean
history to the case of fluctuations is difficult, the basic point remains.
Finally, in the absence of significant clumping reionization requires
approximately one UV photon per baryon in stars so that we may write
\begin{equation}\label{etaX}
\eta_X\equiv \frac{1}{N_{\rm ion,IGM}}=2.5\times10^{-3} \left(\frac{4000}{N_{\rm ion}}\right)\left(\frac{0.1}{f_{\rm esc}}\right).
\end{equation}
Thus reionization should occur after \lya coupling is significant and the gas has been significantly heated.

We conclude that for astrophysical parameters not too far from our fiducial
ones $f_X\approx f_\alpha \approx1$ and $f_*\approx0.1$, \lya
fluctuations will first become important at $z\approx25$ ($\nu\approx
55\MHz$) and temperature fluctuations at $z\approx 19.5$, although this
latter number is considerably more uncertain.  This defines the minimum
requirement for a low-frequency observatory hoping to probe the cosmology
dominated regime before the first sources affect the 21 cm signal.  For the
mean history, \lya coupling becomes relevant at $z\approx23$ and
$T_K>T_{\rm CMB}$ at $z\lesssim 14$.  These results are somewhat sensitive
to the cosmology used, especially the value of $\sigma_8$, which affects
the collapse fraction.  The location of these landmarks will be important
for attempts to probe the mean 21 cm signal.

\section{Observational Prospects}
\label{sec:observation}
\begin{figure}[htbp]
\begin{center}
\includegraphics[scale=0.4]{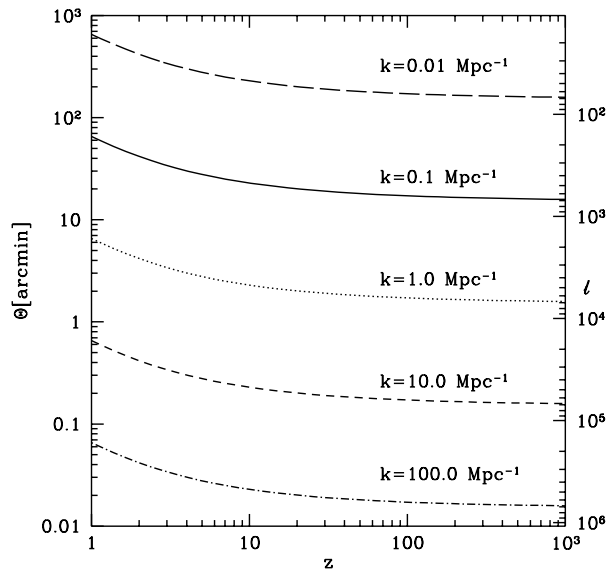}
\caption{Redshift evolution of angular scales.  $\Theta=(2\pi/k)/d_A(z)$, which we map to angular moments using $l\approx\pi/\Theta$.  This is meant to be an approximate guide, rather than an exact conversion.}
\label{fig:angles}
\end{center}
\end{figure}

In section \ref{ssec:f_history}, we showed that detecting the 21 cm signal
would require foreground removal to the level of $10^{-5}$ for reionization
and cosmic twilight and $10^{-7}$ to reach the {\it Dark Ages}.  Next we
explore the experimental capabilities necessary to achieve the required
sensitivity for an accurate measurement of the 21cm power spectrum.  Before
presenting our calculations, let us digress and consider the angular scales
probed by these experiments.  Until this point our calculations have been
presented in terms of comoving wavenumber $k$, appropriate for a 3D power
spectrum. For observers the more natural quantity is the angular scale on
the sky, $\Theta$.  Figure \ref{fig:angles} shows a conversion between the
two using $\Theta=(2\pi/k)/d_A(z)$, with $d_A(z)$ the angular diameter
distance to redshift $z$.  We also relate angles to multipole index $\ell$
in the spherical harmonics decomposition of the sky, based on the
approximate relation $l\approx\pi/\Theta$.  Although this conversion is
somewhat crude, it allows a quick comparison with quantities that might be
more intuitive to the CMB researchers.  Astrophysics primarily affects the
21 cm power spectrum on comoving scales $k=0.1-1 \iMpc$ corresponding to
angular scales of several arcminutes.

We now calculate the signal-to-noise ratio ($S/N=\sqrt{P_{21}/\delta
P_{21}}$) for three fiducial experiments: {\it (i)} a pathfinder class
experiment like the {\it Murchison Widefield Array} (MWA); {\it (ii)} a
fully fledged EoR instrument like the {\it Square Kilometer Array} (SKA);
and {\it (iii)} a futuristic lunar array (LA) \citep{Carilli}.  The
parameters used to describe these instruments are summarized in Table
\ref{tab:telescopes}.  In each case, we specify the number of antennae
$N_a$, and the total collecting area $A_{\rm tot}=N_a A_e$ for an
experiment optimized to observe $z=8$.  We also specify the bandwidth $B$
and total observing time $t_{\rm int}$.  Note that while we have chosen
parameters that correspond roughly to proposed experiments, we take
liberties with the antennae configurations to maximize the sensitivity of
each experiment at each redshift.  We are interested in the most favorable
capabilities of each experiment given the parameters in Table 1, and so we
optimize the associated array configuration for each redshift window.  For
ease of reference, we will label these experiments MWA, SKA, and LA.
Current designs for the actual MWA and SKA limit their frequency range
($\gtrsim 80\,{\rm MHz}$) and antennae distribution, and so the above
labels should not be associated with the actual instruments being designed
or constructed. The labels are simply meant to denote different scales of
experimental effort.

\begin{table}[htdp]
\caption{Low-frequency radio telescopes and their parameters.  We specify the number of antennae $N_a$, total collecting area $A_{\rm tot}$, bandwidth $B$, and total integration time $t_{\rm int}$ for each instrument.}
\begin{center}
\begin{tabular}{ccccc}
\hline
\hline
Array & $N_a$ & $A_{\rm tot}(10^3\,{\rm m^2})$ & $B$ (MHz) & $t_{\rm int}$ (hr)\\
\hline
MWA & 500 & 7.0  & 6 & 1000\\
SKA & 5000 & 600  & 6 & 1000\\
LA & 7800 & 3600 & 8 & 12000\\
\hline
\hline
\end{tabular}
\end{center}
\label{tab:telescopes}
\end{table}%

The variance of a 21 cm power spectrum estimate for a single
$\mathbf{k}$-mode with line of sight component $k_{||}=\mu k$ is given by
\citep{lidz2007}:
\begin{multline}
\sigma_P^2(k,\mu)=\\ \frac{1}{N_{\rm field}}\left[\bar{T}_b^2P_{21}(k,\mu)+T_{\rm sys}^2\frac{1}{B t_{\rm int}}\frac{D^2\Delta D}{n(k_\perp)}\left(\frac{\lambda^2}{A_e}\right)^2\right]^2.
\end{multline}
We restrict our attention to modes in the upper-half plane of the
wavevector ${\bf k}$, and include both sample variance and thermal detector
noise assuming Gaussian statistics. The first term on the right-hand-side
of the above expression provides the contribution from sample variance,
while the second describes the thermal noise of the radio telescope.  The
thermal noise depends upon the system temperature $T_{\rm sys}$, the survey
bandwidth $B$, the total observing time $t_{\rm int}$, the conformal
distance $D(z)$ to the center of the survey at redshift $z$, the depth of
the survey $\Delta D$, the observed wavelength $\lambda$, and the effective
collecting area of each antennae tile $A_e$.  The effect of the
configuration of the antennae is encoded in the number density of baselines
$n_\perp(k)$ that observe a mode with transverse wavenumber $k_\perp$
\citep{mcquinn2005}.  Observing a number of fields $N_{\rm field}$ further
reduces the variance.

Estimates of the error on a power spectrum measurement are calculated using the Fisher matrix formalism, so that the $1-\sigma$ errors on the model parameter $\lambda_i$ are $(\mathbf{F}_{ij}^{-1})^{1/2}$, where 
\begin{equation}
F_{ij}=\sum_{\rm \mu} \frac{\epsilon k^3 V_{\rm survey}}{4\pi^2}\frac{1}{\sigma_P^2(k,\mu)}\frac{\partial P_{T_b}}{\partial \lambda_i}\frac{\partial P_{T_b}}{\partial \lambda_j}.
\end{equation}
In this equation, $V_{\rm survey}=D^2\Delta D(\lambda^2/A_e)$ denotes the
effective survey volume of our radio telescopes and we assume wavenumber
bins of width $\Delta k=\epsilon k$.  We will be interested in the cases
where $\lambda_i=\{\bar{P}_{T_b}\}$ and $\lambda_i=\{
P_{\mu^0},\,P_{\mu^2},\,P_{\mu^4}\}$.

Instrumental sensitivities have been shown to depend upon the distribution
of antennae.  One typical model is to assume a close packed core with
filling fraction close to unity surrounded by antennae distributed with a
$r^{-2}$ dependence out to a sharp cutoff at the edge of the array.
\citet{lidz2007} showed that significant gains in sensitivity can be
achieved by using a ``super core" configuration, condensing the array so
that the filling fraction is unity throughout the array.  Since most of the
longest baselines in the $r^{-2}$ configuration only poorly sample the
largest $k$ modes anyway, this enhances the signal-to-noise while losing
measurements on only a few large $k$ modes.  Since we wish to maximize the
sensitivity of our instrument, we will use a similar ``super core"
configuration.  The effective collecting area of the antennae is a strong
function of observed frequency (scaling as $\nu^{-2}$ or $\lambda^2$), so
long as antennae are more widely spaced than $\lambda/2$.  We optimize our
arrays at each redshift by setting the minimum baseline to $\lambda/2$ and
then close packing the antennae.  Our calculations should then give a
reasonable idea of the best performance that can be hoped for given our
chosen specifications.  In practice, an array will be optimized for a
single redshift and will suffer performance degradation when probing higher
redshifts, where geometrical shadowing becomes important.  Additionally,
some fraction of antennae will be reserved to provide the long baselines
needed for foreground removal.

\begin{figure}[htbp]
\begin{center}
\includegraphics[scale=0.4]{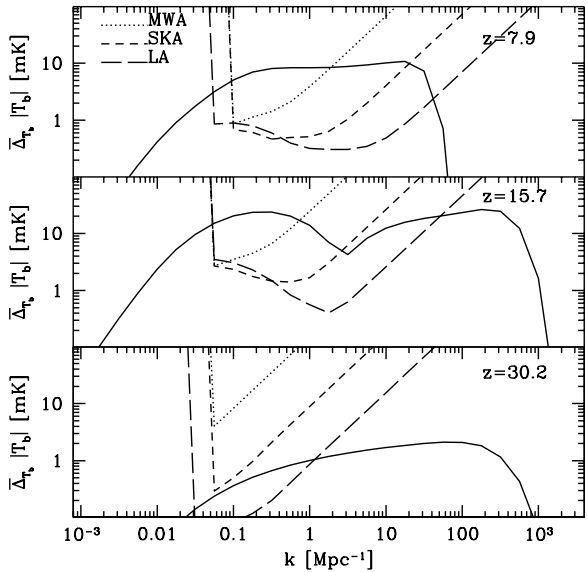}
\caption{Errors on the angle-averaged power spectrum $\bar{\Delta}_{T_b}$
(solid curves) for our optimized experiments MWA (dotted curve), SKA (short dashed curve), and LA
(long dashed curve), as functions of wavenumber $k$ at redshifts $z=7.9$ (top panel),
15.7 (middle panel), and 30.2 (bottom panel).}
\label{fig:errorsK}
\end{center}
\end{figure}

Figure \ref{fig:errorsK} shows three redshift slices of the 21 cm power
spectrum for model A.  These illustrate the shape of the sensitivity of the
various instruments.  The sensitivity is degraded by foregrounds on large
scales and thermal noise on small scales.  In each case, those scales
closest to the foreground cutoff (which is determined by the line-of-sight
resolution set by the experimental bandwidth) on large scales are the most
likely to retain some foreground contamination.  Unfortunately, these tend
to be the modes that are measured with the highest sensitivity.  The shape
of the power spectrum on intermediate scales shows significant evolution as
different sources of fluctuations become important.  At $z=30.2$,
$\bar{\Delta}_{T_b}$ is dominated by density fluctuations.  At $z=15.7$,
temperature fluctuations lead to a trough on a scale $k\approx1\iMpc$ and a
peak at $k\approx0.1\iMpc$.  The location of this trough evolves at higher
redshift to lower $k$, where it might be more easily detectable. By
$z=8$, ionization fluctuations flatten the power spectrum significantly.
The evolution of the shape of the power spectrum during reionization has
been examined in more detail by \citet{lidz2007}, who explored the ability
of the MWA to measure the slope and amplitude of the power spectrum.  While
the SKA is a significant improvement over the MWA, it takes the futuristic
LA to directly observe the trough in $\bar{\Delta}_{T_b}$ at $z=15.7$ or to
detect fluctuations at $z=30.2$.

\begin{figure}[htbp]
\begin{center}
\includegraphics[scale=0.4]{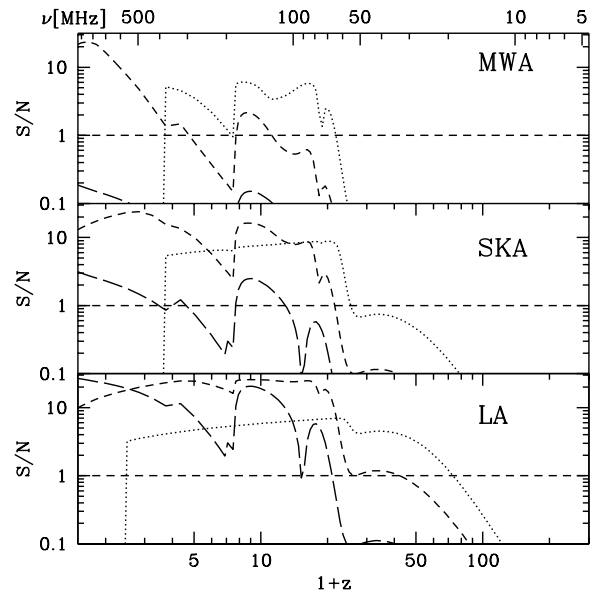}
\caption{Signal to noise ratio as a function of redshift for our optimized experiments.  Curves indicate $k=0.1$ (dotted curve), 1.0 (short dashed curve), and 10$\iMpc$ (long dashed curve). {\em Top panel: } MWA. {\em Middle panel: }SKA. {Bottom panel: }LA.}
\label{fig:errorsSN}
\end{center}
\end{figure}
Figure \ref{fig:errorsSN} shows achievable S/N ratio for measurements of
the spherically averaged signal $\bar{\Delta}_{T_b}$ for the three
instruments as a function of redshift for three different wavenumbers
$k=0.1$, 1, and 10 $\iMpc$.  These wavenumbers span the scales to which
these instruments are sensitive.  We see that a well optimized MWA achieves
high signal-to-noise ratio $S/N$ for only the $k=0.1\iMpc$ mode and
moderate S/N for $k=1.0\iMpc$.  SKA achieves the sample variance limit at
$k=0.1\iMpc$ over a wide range of scales and shows sensitivity to $k=1.0
\iMpc$ through most of the EoR.  Only the LA is capable of detecting the
signal from the {\it Dark Ages} at $z>30$.  These curves are optimistic
since the instruments are assumed to be optimized to the specific frequency
of observation.  At very low redshifts ($z\lesssim3$), the bandwidth that
we have chosen is inadequate to the task of removing foregrounds on the
scale $k=0.1\iMpc$, hence the sharp cutoff in sensitivity seen in Figure
\ref{fig:errorsSN}.

\begin{figure}[htbp]
\begin{center}
\includegraphics[scale=0.4]{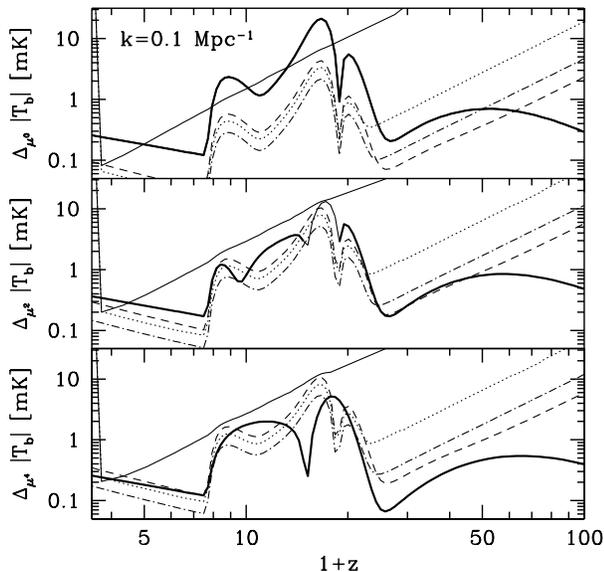}
\caption{Experimental sensitivity to the separation of powers.  Plotted is
the redshift evolution of the power spectrum (thick solid curve) at
$k=0.1\iMpc$ and the corresponding sensitivity for MWA (thin solid curve),
SKA (thin dotted curve), and LA (thin dashed curve).  We also plot
sensitivity for LA with its observing time split between 16 separate fields
(thin dot-dashed curve). {\em Top panel: } $\Delta_{\mu^0}$. {\em Middle
panel: } $\Delta_{\mu^2}$. {\em Bottom panel: } $\Delta_{\mu^4}$.
Calculations are for model A.}
\label{fig:errorsMu}
\end{center}
\end{figure}

We now turn to the possibility of performing the angular separation by
powers of $\mu$ discussed earlier.  There are two main reasons for
performing this separation.  First, as discussed in section
\S\ref{sec:results} the 21 cm fluctuations arising from different
astrophysical models tend to overlap, giving $\bar{\Delta}_{T_b}$ a
complicated shape.  In order to constrain astrophysics, it may be
advantageous to turn to the $\Delta_{\mu^2}$ term.  This term shows more
structure than the spherically averaged signal (see Figure
\ref{fig:errorsMu}), while being simpler to analyze as it is composed of
fewer terms.  Second, the complexity of the astrophysical fluctuations
obscures the underlying cosmological information contained in the density
contribution to 21 cm fluctuations.  This problem can be circumvented by
measuring $\Delta_{\mu^4}$, which is determined by the density field alone
\citep{bl2005sep}.  Although the use of this term is most important during
the astrophysics dominated regime, it might also be important during the
{\it Dark Ages}.  If exotic energy injection mechanisms, such as decaying
dark matter \citep{furlanetto2006dm}, occur during the {\it Dark Ages} they
would produce fluctuations in temperature and ionization fraction that
would interfere with the extraction of cosmology from the density
fluctuations.  The angular separation would provide extra information about
these processes and still allow cosmological constraints to be extracted.

Figure \ref{fig:errorsMu} shows the sensitivity of our fiducial experiments
to the angular components as a function of redshift for wavenumber
$k=0.1\iMpc$.  All three experiments show some sensitivity to
$\Delta_{\mu^0}$, which is to be expected from our analysis of the
spherically averaged signal.  Only SKA and LA show any sensitivity to $\Delta_{\mu^2}$ and $\Delta_{\mu^4}$.  

Since $\Delta_{\mu^0}$ dominates the signal, the errors in $\Delta_{\mu^2}$
and $\Delta_{\mu^4}$ track its shape.  This results in only a few windows
where observation of these components is possible.  During the end of
reionization the largely isotropic ionization signal dominates
\citep{mcquinn2005} making observations of the $\mu^2$ and $\mu^4$
components difficult.  This is also true earlier when temperature and \lya
fluctuations dominate the signal.  Observational prospects for the angular
separation are best when these fluctuations are small, at the beginning of
the emission period but before reionization is well advanced ($8\lesssim
z\lesssim 12$ for model A).  There is also a possibility for measuring
$\Delta_{\mu^2}$ during the absorption period ($15\lesssim z\lesssim 21$
for model A) although this relies somewhat upon the scale-dependent
cancellation between temperature and \lya fluctuations.  We note that a
high $S/N$ detection is required for the angular separation to be feasible.
For this reason, none of the instruments are able to perform the angular
separation during the {\it Dark Ages}.  The angular decomposition also
becomes more difficult on small scales.

Since the SKA and LA experiments are essentially sample variance limited at
$k=0.1\iMpc$, it would seem that performing the full angular separation
will be exceedingly difficult except in a few redshift windows.  This is a
direct consequence of the isotropic part of the power spectrum dominating
the uncertainty in measurement of the $\mu^2$ and $\mu^4$ terms.  To
improve on the results illustrated here it would be necessary to beat down
cosmic-variance by observing multiple fields on the sky.  Since the
variance scales as $N_{\rm field}^{-1/2}$ while the contribution from
thermal noise scales as $t_{\rm int}^{-1}$, there is a trade-off involved
in splitting observing time into different fields.  The experiment will
gain sensitivity on scales to which it is sample variance limited while
losing sensitivity where thermal noise is important.  Only if the
experiment has sufficient instantaneous sensitivity, will it be worth
observing multiple fields.  Note though that for some interferometer
designs, e.g. LOFAR, multiple fields can be observed
simultaneously,improving the S/N while adding only to the computational
cost of the required correlations.  For an experiment like LA, which has
the sensitivity to reach sample-variance limited measurements, experimental
design will be key in distributing antennae and observing time to minimize
errors.

Since performing the angular separation may be vital to obtaining both
astrophysical and cosmological information, we briefly consider a modified
version of the LA in which we redistribute the same fixed total observing
time from deep integration of a single field, necessary to probe the {\it
Dark Ages}, into shallower integrations of 16 separate fields.  This is
plotted in Figure \ref{fig:errorsMu} (dot-dashed curve).  We see that
considerable gains in sensitivity are achievable, dramatically increasing
the range of redshifts over which the angular separation is possible at
high S/N on this scale.  However, this version of the LA is unable to probe
the {\it Dark Ages} and is sample variance limited over a smaller range of
wavenumbers.  Since observation strategies for direct observation of the
{\it Dark Ages} versus separating $\Delta_{\mu^4}$ are orthogonal to one
another, detailed consideration of which approach is better for cosmology
will be important for the design of future instruments.

\section{Discussion and conclusions}
\label{sec:conclude}

Future low-frequency radio observations of the redshifted 21 cm line hold
the potential to greatly increase our knowledge of the high redshift
Universe.  Theoretical understanding of fluctuations in the 21 cm
brightness temperature has reached the point where it is possible to begin
making predictions for the evolution of the signal from recombination to
the present day.  In this paper, we have combined the present understanding
of fluctuations in density, neutral fraction, \lya flux, and X-ray heating,
to calculate the 3-dimensional 21 cm power spectrum as a function of
redshift.  This draws attention to the three different epochs mentioned in
the Introduction.  Early reionization compresses the reionization and
twilight signal into a narrow plateau, likely complicating attempts to
separate different fluctuations and learn about astrophysics.

Our comprehensive analysis provides a useful foundation for optimizing the
design of future arrays whose goal is to separate the particle physics from
the astrophysics, either by probing the peculiar velocity distortion of the
21 cm power spectrum during the epoch of reionization, or by extending the
21 cm horizon to $z\gtrsim 20$ or $z\lesssim 6$.

We have shown that the signal above $z=25$ is likely to be relatively
uncontaminated by astrophysics.  This sets the minimum redshift that a
low-frequency radio observatory must probe in order to seek cosmology most
simply.  This same calculation indicates that experiments with sensitivity
down to 60 MHz should be able to observe the full span of the EoR signal.

Observations should be sensitive to both the emission and absorption
regimes of the 21 cm signal during the EoR.  This is in part due to
the greater amplitude of $T_b$ during the absorption regime and in
part to the enhancement of the signal from temperature and \lya
fluctuations.  Additionally, experiments sensitive to $z<5$ should
have a chance of detecting the post-reionization signal.

In order for experiments to detect 21 cm fluctuations from the EoR they
will need to first remove foregrounds to the level of $10^{-5}$.  To probe
the {\it Dark Ages}, heroic efforts to obtain the level of $10^{-7}$ will
be required.  If this can be achieved then the SKA should be able to
measure the EoR signal over several decades of wavenumber.  Significant
improvement in either collecting area or, for example, going to the moon
for longer integration times will be necessary to access the {\it Dark
Ages}.

Interestingly, the 21 cm emission from residual hydrogen after reionization
($z\lesssim 6$) offers excellent prospects for probing fundamental physics
\citep{wyithe2007}, because of the steep decline in the Galactic foreground
brightness with decreasing wavelength ($\propto \lambda^{2.6}$).  The
pockets of self-shielded hydrogen are expected to trace the distribution of
matter, and allow a precise determination of the matter power
spectrum. This in turn will provide an unprecedented probe of
non-Gaussianity and running of the spectral index of the power-spectrum of
primordial density fluctuations from inflation. Detection of small-scale
fluctuations can also be used to infer the existence of massive neutrinos
\citep{WL08} and other sub-dominant components in addition to the commonly
inferred cold dark matter particles. The scale of the baryonic acoustic
oscillations in the power spectrum can be used as a standard ruler to
constrain the equation of state of the dark energy through 99\% of the
cosmic history \citep{BLBAO,wyithe2007bao,Pen}.

The use of 21cm cosmology to study fundamental physics is not restricted to
either high redshifts ($z\gtrsim 25$; $\nu\lesssim 50$MHz) or low redshifts
($z\lesssim 6$; $\nu\gtrsim 200$MHz). The implications for
inflation and dark matter can be separated from the astrophysical effects
of star formation through an angular decomposition of the 21cm power
spectrum, even during reionization ($6\lesssim z\lesssim 25$). The angular
term $\Delta_{\mu^4}$ isolates the contribution of gravitationally induced
peculiar velocities. We have found that with an appropriately designed
instrument, separating the $\Delta_{\mu^4}$ term should be possible over a
wide range of redshifts.

\acknowledgments 
 
JRP is supported by NASA through Hubble Fellowship grant HST-HF-01211.01-A
awarded by the Space Telescope Science Institute, which is operated by the
Association of Universities for Research in Astronomy, Inc., for NASA,
under contract NAS 5-26555.


 
 \end{document}